\documentclass[runningheads]{llncs}
\usepackage[T1]{fontenc}
\usepackage{graphicx}
\usepackage{amsmath}
\usepackage{tabularx,booktabs}
\usepackage{subfigure}
\usepackage{hyperref}
\usepackage{float}

\usepackage{graphicx}
\usepackage{subcaption}
\usepackage{hyperref}
\usepackage{float}

\usepackage{booktabs}

\usepackage{tabularx} 
\usepackage{mathtools, nccmath}

\begin{document}
\title{Automated Segmentation and Analysis of \\ Cone Photoreceptors in Multimodal Adaptive Optics Imaging}
\author{Prajol Shrestha\inst{1} \and
Mikhail Kulyabin\inst{1} \and
Aline Sindel\inst{1} \and \\
Hilde R. Pedersen\inst{2} \and 
Stuart Gilson\inst{2} \and \\
Rigmor Baraas\inst{2} \and
Andreas Maier\inst{1}}
\authorrunning{Shrestha et al.}
\institute{Pattern Recognition Lab, Department of Computer Science, \\ Friedrich-Alexander-Universität Erlangen-Nürnberg, Erlangen, Germany \\ \email{prajol.shrestha@fau.de}\\ \and
National Centre for Optics, Vision and Eye Care, \\ Faculty of Health and Social Sciences, \\ University of South-Eastern Norway, Kongsberg, Norway}

\titlerunning{Automated Segmentation and Analysis of Cone Photoreceptors}

\maketitle              % typeset the header of the contribution
\begin{abstract}
Accurate detection and segmentation of cone cells in the retina are essential for diagnosing and managing retinal diseases. In this study, we used advanced imaging techniques, including confocal and non-confocal split detector images from adaptive optics scanning light ophthalmoscopy (AOSLO), to analyze photoreceptors for improved accuracy. Precise segmentation is crucial for understanding each cone cell's shape, area, and distribution. It helps to estimate the surrounding areas occupied by rods, which allows the calculation of the density of cone photoreceptors in the area of interest. In turn, density is critical for evaluating overall retinal health and functionality. We explored two U-Net-based segmentation models: StarDist for confocal and Cellpose for calculated modalities. Analyzing cone cells in images from two modalities and achieving consistent results demonstrates the study's reliability and potential for clinical application.

\end{abstract}
\section{Introduction}
Cone and rod photoreceptors are the fundamental units of the retina responsible for detecting light. Cones are sensitive to bright light and responsible for color perception. They are densely concentrated in the central fovea within the macula, with their density decreasing rapidly towards the periphery. Rods are sensitive to low light and enable night vision. They are predominantly located in the peripheral regions of the retina~\cite{kolb1995webvision}. The distribution of photoreceptors varies with eccentricity, allowing the eye to function effectively in different lighting conditions and perceive fine details and a wide range of colors.

Retinal diseases such as age-related macular degeneration, Stargardt disease, and diabetic retinopathy pose significant threats to vision, particularly when they affect cone cells, which are crucial for central and color vision~\cite{10.3389/fopht.2024.1384473}. If these conditions are not diagnosed early, they can lead to severe and often irreversible vision loss. Early detection is essential for effective treatment, allowing interventions that can slow the disease's progression or mitigate its impact on vision. However, detecting these diseases in their early stages is challenging due to the subtle nature of early signs and symptoms and limitations in imaging technologies.

Advanced imaging techniques like adaptive optics scanning light ophthalmoscopy (AOSLO) have become invaluable in addressing retinal diseases. AOSLO provides high-resolution images of the individual photoreceptor cells and supports various modalities, including confocal, non-confocal, and darkfield~\cite{Roorda2015}. The confocal modality reveals the centers of cone photoreceptors by imaging the functional cone outer segment structure, while the non-confocal modality, most commonly the split detector, reveals rod cells and cone boundaries through the inner segment structure. The darkfield modality highlights the retinal pigment epithelium (RPE) cells, providing additional structural information~\cite{10.1167/iovs.63.2.8}. The combination of these modalities offers complementary views of the retinal structure, enhancing our ability to study various retinal diseases and conditions.

Conventional method presented in work by Li and Roorda~\cite{li2007automated} rely on the optical fiber properties of cone photoreceptors and can mislabel rods as cones. Therefore, it needs to be revised by a human expert. Creating a fully automatic method for segmenting and detecting cone and rod areas will significantly increase the possibilities of retinal research. 

Recently, Kulyabin et al. introduced a method based on deep learning (DL) for automated detection and segmentation of cones in AOSLO images, using the Voronoi algorithm to generate the photoreceptor outlines~\cite{kulyabin2024generalist}. However, the method is limited to the range of degrees of eccentricity from the foveal center — $0^{\circ}$, $1^{\circ}$, and $2^{\circ}$. As the size of rods, and consequently rods total area, increases, approximation using the Voronoi algorithm, which uses the centers of cones to generate outlines, becomes less accurate. 

In this work, to independently define zones of cones and rods, we investigated different methods for annotating cones for the further training of DL models and calculating photoreceptor densities for a range of $0^{\circ}$ - $10^{\circ}$ of eccentricity from the fovea. Precise segmentation is crucial as it enables a detailed understanding of the shape and area of each cone and helps estimate the areas occupied by rods. U-Net-based methods such as StarDist~\cite{schmidt2018cell} and Cellpose~\cite{pachitariu2022cellpose} showed strong performance in the segmentation confocal and non-confocal split detection (referred to as calculated below as the image used for analysis is the difference between the signals from the non-confocal detectors, divided by their sum) images~\cite{scoles2014vivo}, respectively. Furthermore, the precise analysis of cones from both modalities improves the ability to detect abnormalities with higher precision, thereby supporting each other's findings with more informed clinical decisions.

\bigbreak
\bigbreak

\section{Materials and Methods}
%\subsection{Dataset}
%data
In this study, we utilized a semi-automatically labeled dataset of AOSLO images of 11 participants with normal vision, covering an age range of 14 to 65. This diverse sample represents a broad spectrum of healthy retinas, providing valuable insights into retinal structure across different stages of life. We explored various annotation techniques, such as manual annotation and a human-in-the-loop method. After initial manual annotation, we further processed the data by separating regions based on each cone cell's center information using the Voronoi method or using boundary vertices. Additionally, we transformed polygonal annotations into circular shapes using the closest vertex and closest ridge midpoint vertex methods to standardize the representation of cone cells. In the human-in-the-loop, annotations are automatically predicted, and a human expert makes the necessary corrections.

The dataset was split at the participant level, with 8 participants used for training (272 confocal and 20 calculated images) and 3 (96 confocal and 86 calculated images) for testing. Due to suboptimal performance simultaneously for all eccentricities, images were further divided into central fovea ($0^{\circ}$ -- $4^{\circ}$) and parafovea ($5^{\circ}$ -- $10^{\circ}$) groups, with further independent training of the models for each of the groups.

\begin{figure*}[h]
    \centering
    \includegraphics[width=1\textwidth]{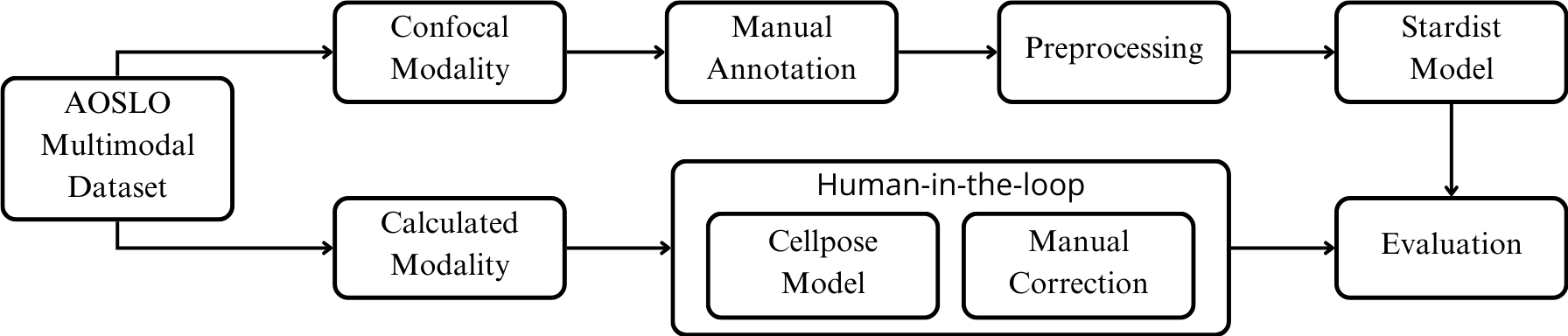}
    \caption{Pipeline for processing multimodal AOSLO dataset, incorporating images of confocal and calculated modalities with manual and semi-automated annotation steps.}
    \label{fig:pipeline}
\end{figure*}

Fig.~\ref{fig:pipeline} shows the proposed pipeline. For the detection of cone cells in images from confocal modality, we fine-tuned the StarDist model~\cite{schmidt2018cell}, which is based on the original U-Net architecture, utilizing a star-convex polygon for precise localization. To extract the precise shape and location of a cone cell, StarDist evaluates the likelihood of a pixel being the cell center through a predicted probability map and utilizes the predicted distance map to calculate the distance from the center to the cell's boundary. Similarly, for images from calculated modality, we fine-tuned the Cellpose model~\cite{10.1038/s41592-020-01018-x}, which is based on a U-Net with residual blocks and an image style transfer. The Cellpose model generates a vector flow field using the predicted horizontal and vertical gradients of the image, and by tracking this field, pixels belonging to each cone cell are grouped and further refined with the predicted binary map to extract the exact shapes of the cells.

\section{Results}
The StarDist model was trained for 600 epochs with 100 steps per epoch, using ADAM optimizer with a learning rate of 0.0003, a batch size of 4, and a patch size of 64 $\times$ 64 pixels. We used the Cellpose GUI to annotate automatically and adjust the results. The training involved 5 to 10 calculated images per group, with each image trained over 200 epochs using stochastic gradient descent, a learning rate of 0.1, and a weight decay of 0.0001. Both models were trained on a single NVIDIA RTX 2080Ti GPU. IoU and the Dice score were used to evaluate the segmentation performance, ensuring alignment between predicted outputs and actual cell boundaries. 

\begin{figure*}[h]

    \centering
    \includegraphics[width=1\textwidth]{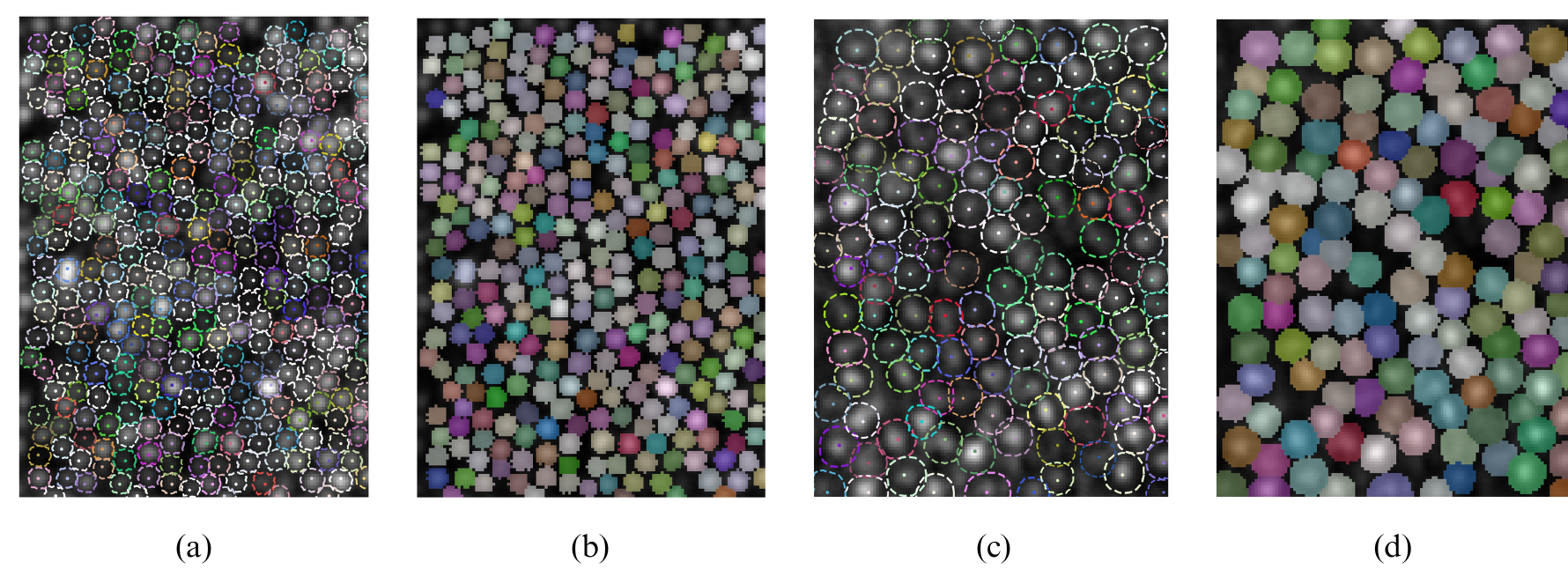}

     \caption{Visualization of StarDist model predictions on confocal images: cone predictions in central fovea: centers and boundaries (a), segmentation masks (b); cone predictions in parafovea: centers and boundaries (c), segmentation masks (d).}
      \label{fig:stardist_pred}
\end{figure*}

\begin{figure*}[h]

    \centering
    \includegraphics[width=0.8\textwidth]{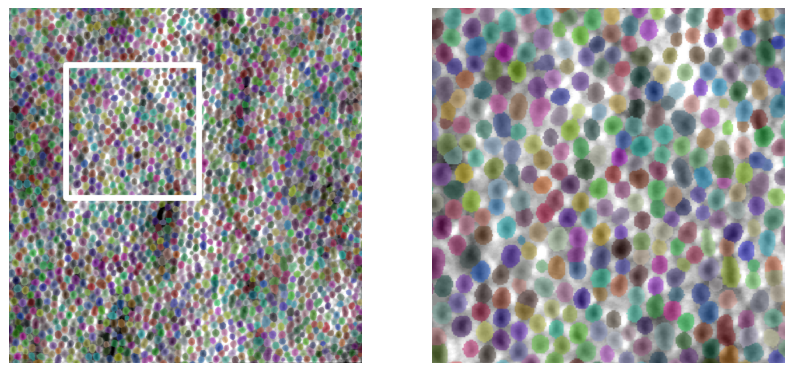}

     \caption{Visualization of prediction performance on the parafovea calculated image using the Cellpose 2.0 model. The white box highlights the zoomed area in the right column.}
      \label{fig:cellpose_pred}
\end{figure*}

Fig.~\ref{fig:stardist_pred} presents the predictions on confocal images using the StarDist model, with predictions for central fovea and parafovea. Fig.~\ref{fig:cellpose_pred} displays parafovea predictions on calculated images using the Cellpose model. These predictions appear promising, with only minimal errors, indicating the strong performance of both models in segmenting and detecting the cells.

% \begin{figure*}[h]
% \begin{minipage}[c]{0.6\textwidth}
% \includegraphics[width=\textwidth]{output_image.png}
% \end{minipage}\hfill
% \begin{minipage}[c]{0.35\textwidth}
% \caption{
%        Visualization of prediction performance on the parafovea calculated image using the Cellpose 2.0 model. The white box highlights the zoomed area in the right column.
%     }
% \label{fig:cellpose_pred}
% \end{minipage}
% \end{figure*}

\begin{table}[]
\setlength{\tabcolsep}{5 pt}
    \centering
    \caption{Evaluation metrics for segmentation models on two modalities}
    \begin{tabular}{@{}llcccccc@{}}
        \toprule
        Modality & Annotation Type      &    \multicolumn{6}{c}{Eccentricity} \\
            &     & \multicolumn{2}{c}{Central fovea} & \multicolumn{2}{c}{Parafovea} & \multicolumn{2}{c}{All} \\ 
                 &                          & IoU         & Dice       & IoU         & Dice       & IoU         & Dice \\ \midrule
        Calculated & Human-in-the-loop       & \textbf{0.52} & \textbf{0.68} & \textbf{0.54} & \textbf{0.70} & \textbf{0.50} & \textbf{0.67} \\
        Confocal   & Closest Vertex          & \textbf{0.67} & \textbf{0.80} & \textbf{0.78} & \textbf{0.88} & \textbf{0.69} & \textbf{0.81} \\ 
        Confocal   & Closest Ridge Midpoint  & 0.48          & 0.61          & 0.58          & 0.70          & 0.53          & 0.64 \\ 
        Confocal   & Manual                  & 0.61          & 0.74          & 0.69          & 0.81          & 0.63          & 0.86 \\

        \bottomrule
    \end{tabular}
    \label{tab:metrics}
\end{table}

\begin{figure*}[htbp]
     \centering
     \includegraphics[height=6.5cm]{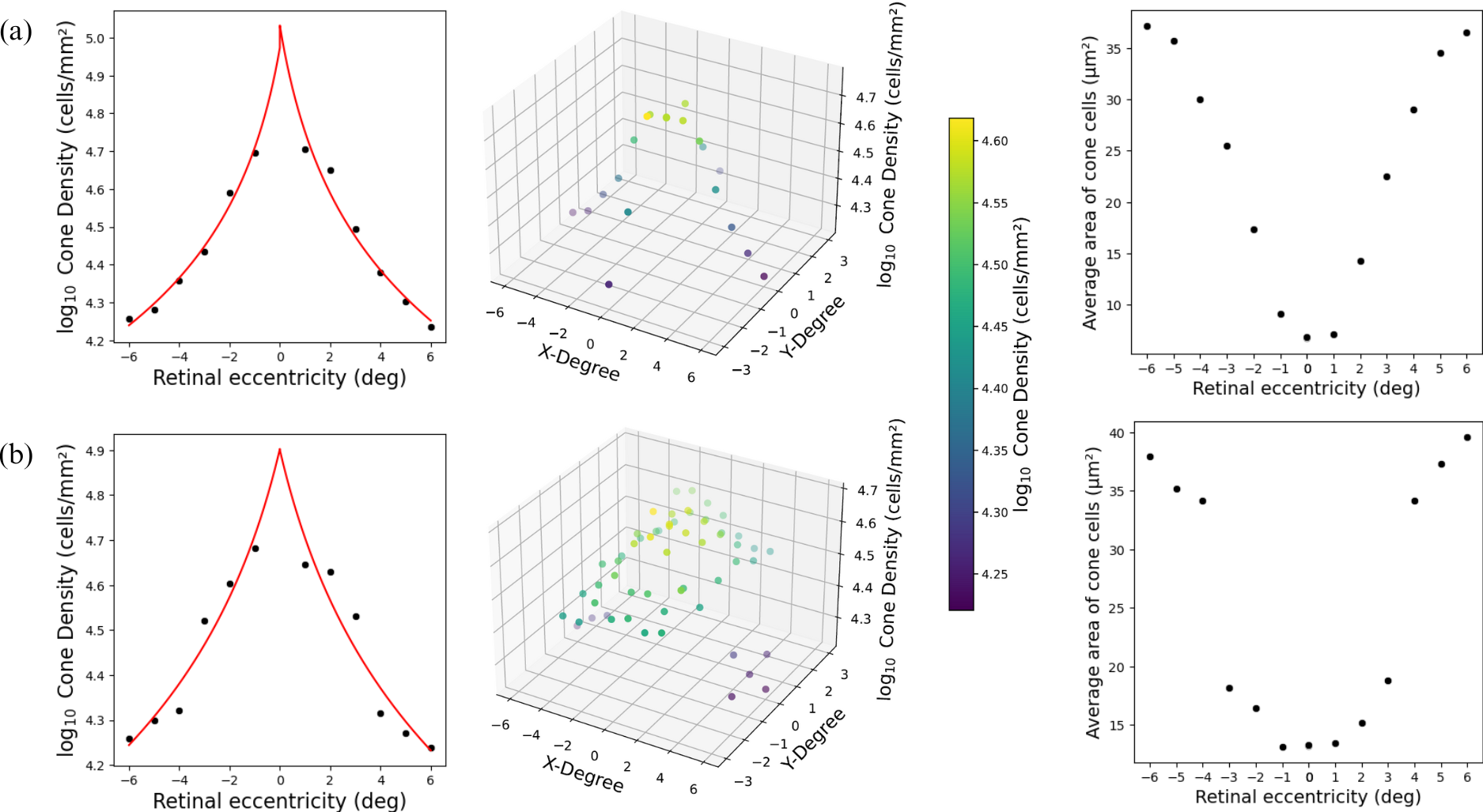}
      \caption{Visualizations of cone density and average cone area as a function of eccentricity: cone density fitted with asymmetric power function, volumetric cone density, average cone area computed using confocal modality (a) and calculated modality (b) images from the same participant.}
  \label{fig:density_using_confocal_and_calculated}
 \end{figure*}

Table~\ref{tab:metrics} shows the experiment results. The prediction results from both models were evaluated using the IoU and Dice metrics. For confocal images with the closest vertex annotation style, the central fovea segmentation achieved an IoU of 0.67 and a Dice score of 0.80, while parafovea segmentation performed better, with an IoU of 0.78 and a Dice score of 0.88. In contrast, the model performed less effectively with other annotation styles. For calculated images, the results were lower, with IoU of 0.52 and a Dice score of 0.68 for the central fovea and IoU of 0.54 and a Dice score of 0.70 for parafovea. Additionally, there was a strong correlation between the predicted and ground truth masks, as confirmed by a Pearson correlation coefficient of 0.850 and a Spearman correlation coefficient of 0.847. To compute the density of cones (\(d\)) and the mean cone area (\(\bar{A}\)) according to predicted cone centers and mask, we applied Eq.~\ref{eq:density} and Eq.~\ref{eq:mean_area}:

\begin{equation}
d = \frac{N}{h \times w \times \left(\frac{\mu}{s_{\!f}}\right)^2},
\label{eq:density}
\end{equation}

\begin{equation}
\bar{A} = \frac{\sum (p \times \mu^2)}{N},
\label{eq:mean_area}
\end{equation}

\noindent
where \(N\) is the total number of cones, \(h\) and  \(w\) are the height and the width of the area in pixels, \(\mu\) denotes the spatial resolution (microns per pixel), \(s_{\!f}\) is the scaling factor, and \(p\) represents the pixel count per cone. The density values are fitted (Fig. ~\ref{fig:density_using_confocal_and_calculated}) according to the asymmetric power function ~\cite{10.1167/iovs.63.2.8}:

% Asymmetric power function
\begin{equation}
\log(d_i) = \kappa + k_s + 
\left\{\!\begin{aligned}
&(\pi_n + p_s) \log(|r_i| + (\rho + r_s)), &\text{ if } r_i \geq 0 \\[1ex]
&(\pi_n - \pi_t + p_{ns} - p_{ts}) \log(\rho + r_s) + \\
&(\pi_t + p_{ts}) \log(|r_i| + (\rho + r_s)), &\text{ if } r_i < 0
\end{aligned}\right\} + \varepsilon_i,
\label{eq:power_fun}
\end{equation}

\noindent
where \(d_i\) is the predicted density at eccentricity \(r_i\) (with negative values for nasal), \(\pi_n\) is the fixed-effect exponent for the nasal visual field, \(\pi_t\) is the fixed-effect exponent for the temporal visual field, \(\varepsilon_i\) denotes the random error of the \(i\)-th observation, \(p_s\) is the participant-specific fixed effect, \(p_{ns}\) is the random participant error for the nasal fixed effect, and \(p_{ts}\) is the random participant error for the temporal fixed effect. Furthermore, the random errors, random intercept \( k_s \), and participant-specific effects \( p_{ns} \), \( p_{ts} \), and \( r_s \) are each drawn from normal distributions with the mean of zero and their respective variances \( \sigma^2 \), \( \sigma_s^2 \), \( \sigma_{ns}^2 \), \( \sigma_{ts}^2 \), and \( \sigma_{rs}^2 \). 

Fig.~\ref{fig:density_using_confocal_and_calculated} highlights the distribution of cone inner segment areas as a function of eccentricity, providing a clear visualization of the variation in average cone sizes. This analysis is crucial for understanding the spatial organization of the retina by isolating cones from rods, resulting in a more precise evaluation of cone photoreceptors and contributing to a more accurate assessment of retinal health.

\section{Discussion}
We proposed an automated approach to segment cone photoreceptors and measure their density and inner segment area as a function of eccentricity in multimodal AOSLO images, isolating cones from rods using DL models for a detailed understanding of retinal structure. Our findings align with earlier studies on how cone density changes with distance from the fovea. Additionally, we observed that the cones' average inner segment area is smaller near the fovea, where the cone density is highest, and gradually increases as we move further away~\cite{scoles2014vivo}. This trend is significant, as it provides insights into the spatial organization of cones, which can be helpful in conditions like Stargardt disease. We generated detailed cone density maps, allowing us to analyze how cone distribution relates to other retinal features. The models perform well in segmenting cone boundaries. However, detecting cones in low-contrast areas or where cones are tightly packed is more challenging, particularly within the central fovea. Variations in image quality, such as noise or motion blur, can also affect the results. As with many automated methods, improving the prediction is challenging, particularly when the boundaries are unclear, leading to differences in ground truth generation. To address this, we are using active learning with human feedback to increase the quality and consistency of ground truth and predictions. Furthermore, integrating novel SOTA models such as SAM (Segment Anything Model) is a potential improvement in handling more complex issues.

\bibliographystyle{splncs04}
\bibliography{refs}

\begin{thebibliography}{10}
\providecommand{\url}[1]{\texttt{#1}}
\providecommand{\urlprefix}{URL }
\providecommand{\doi}[1]{https://doi.org/#1}

\bibitem{10.1167/iovs.63.2.8}
Baraas, R.C., Pedersen, H.R., Knoblauch, K., Gilson, S.J.: {Human Foveal Cone and RPE Cell Topographies and Their Correspondence With Foveal Shape}. Investigative Ophthalmology and Visual Science  \textbf{63}(2), ~8--8 (02 2022)

\bibitem{kolb1995webvision}
Kolb, H., Fernandez, E., Nelson, R.: Webvision: The Organization of the Retina and Visual System. University of Utah Health Sciences Center (1995)

\bibitem{kulyabin2024generalist}
Kulyabin, M., Sindel, A., Pedersen, H., Gilson, S., Baraas, R., Maier, A.: Generalist segmentation algorithm for photoreceptors analysis in adaptive optics imaging. arXiv preprint arXiv:2408.14810  (2024)

\bibitem{li2007automated}
Li, K., Roorda, A.: Automated identification of cone photoreceptors in adaptive optics retinal images. JOSA A  \textbf{24}(5),  1358--1363 (2007)

\bibitem{pachitariu2022cellpose}
Pachitariu, M., Stringer, C.: Cellpose 2.0: how to train your own model. Nature Methods  \textbf{19},  1634--1641 (2022)

\bibitem{10.3389/fopht.2024.1384473}
Pedersen, H.R., Gilson, S., Hagen, L.A., Holtan, J.P., Bragadottir, R., Baraas, R.C.: Multimodal in-vivo maps as a tool to characterize retinal structural biomarkers for progression in adult-onset stargardt disease. Frontiers in Ophthalmology  \textbf{4} (2024)

\bibitem{Roorda2015}
Roorda, A., Duncan, J.L.: Adaptive optics ophthalmoscopy. Annual Review of Vision Science  \textbf{1}(1),  19--50 (2015)

\bibitem{schmidt2018cell}
Schmidt, U., Weigert, M., Broaddus, C., Myers, G.: Cell detection with star-convex polygons. In: MICCAI 2018: 21st International Conference, Granada, Spain, September 16-20, 2018, Proceedings, Part II 11. pp. 265--273. Springer (2018)

\bibitem{scoles2014vivo}
Scoles, D., Sulai, Y.N., Langlo, C.S., Fishman, G.A., Curcio, C.A., Carroll, J., Dubra, A.: In vivo imaging of human cone photoreceptor inner segments. Investigative ophthalmology \& visual science  \textbf{55}(7),  4244--4251 (2014)

\bibitem{10.1038/s41592-020-01018-x}
Stringer, C., Wang, T., Michaelos, M., Pachitariu, M.: {Cellpose: a generalist algorithm for cellular segmentation}. Nature Methods  \textbf{18}(1),  100--106 (2021)

\end{thebibliography}
%\end{thebibliography}
\end{document}